\documentclass[preprint,12pt]{elsarticle}
\journal{Neurocomputing}

\usepackage{booktabs}
\usepackage{multirow}
\usepackage{siunitx}
\usepackage{subcaption}
\usepackage{multicol}
\usepackage{url}
\usepackage[export]{adjustbox}
\usepackage{algorithm,algorithmic}
\usepackage{lettrine}
\usepackage [fleqn] {amsmath}
\usepackage{graphicx}
\usepackage[T1]{fontenc}
\usepackage{colortbl}
\usepackage{color}
\usepackage{rotating}
\usepackage{float}
\usepackage{caption}
\usepackage{hhline}
\usepackage{textcomp}
\usepackage{amsfonts}

\usepackage{blindtext}
\usepackage{bm}
\usepackage{xcolor}
\usepackage{amsmath}
\newcommand\amend[1]{\textcolor{blue}}

\usepackage{mathabx}
\usepackage{tabularx}
\usepackage{pifont}
\usepackage{hyperref}
\usepackage{footnote}
\usepackage{listings}
\usepackage{acronym}
\usepackage[htt]{hyphenat}

\acrodef{OSP}[OSP]{Original Software Publication}
\acrodef{RRAM}[RRAM]{Resistive Random-Access Memory}
\acrodef{ADC}[ADC]{Analog to Digital Converter}
\acrodef{DL}[DL]{Deep Learning}
\acrodef{MAC}[MAC]{Multiply Accumulate}
\acrodef{DNN}[DNN]{Deep Neural Network}
\acrodef{CNN}[CNN]{Convolutional Neural Network}
\acrodef{ML}[ML]{Machine Learning}
\acrodef{VMM}[VMM]{Vector-Matrix Multiplication}
\acrodef{SPICE}[SPICE]{Simulation Program with Integrated Circuit Emphasis}
\acrodef{MDNN}[MDNN]{Memristive Deep Neural Network}
\acrodef{MDLS}[MDLS]{Memristive Deep Learning System} 
\acrodef{GPU}[GPU]{Graphics Processing Unit}
\acrodef{CPU}[CPU]{Central Processing Unit}
\acrodef{API}[API]{Application Programming Interface}
\acrodef{0T1R}[0T1R]{0-Transistor 1-Resistor}
\acrodef{1T1R}[1T1R]{1-Transistor 1-Resistor}
\acrodef{ISA}[ISA]{Instruction Set Architecture}
\acrodef{CIM}[CIM]{Compute-In-Memory}
\acrodef{PyPi}[PyPi]{Python Package Index}
\acrodef{WL}[WL]{Word Line}
\acrodef{BL}[BL]{Bit Line}
\acrodef{SL}[SL]{Select Line}
\acrodef{C2C}[C2C]{Cycle-to-cycle}
\acrodef{PCM}[PCM]{Phase Change Memory}
\acrodef{TDM}[TDM]{Time-division Multiplexing}
\acrodef{LUT}[LUT]{Lookup Table}
\acrodef{SGD}[SGD]{Stochastic Gradient Descent}
\acrodef{DSE}[DSE]{Design Space Exploration}

\date{}

\makeatletter
\def\ps@pprintTitle{%
 \let\@oddhead\@empty
 \let\@evenhead\@empty
 \def\@oddfoot{\hbox to \textwidth{\@myfooterfont{Accepted for Publication in Neurocomputing\hfill}}}%
 \let\@evenfoot\@oddfoot}
\makeatother

\begin{document}

\begin{frontmatter}

\title{MemTorch: An Open-source Simulation Framework for Memristive Deep Learning Systems}
\author[JCU]{Corey Lammie}
\author[Latrobe]{Wei Xiang}
\author[IMSE-CNM]{Bernabé Linares-Barranco}
\author[JCU]{Mostafa Rahimi Azghadi\corref{correspondingauthor}}
\address[JCU]{College of Science and Engineering, James Cook University, Australia}
\address[Latrobe]{School of Computing, Engineering and Mathematical Sciences, La Trobe University, Australia}
\address[IMSE-CNM]{Institute of Microelectronics of Seville, IMSE-CNM, Parque Tecnológico de la Cartuja, CSIC, University of Seville, Spain \vspace{-0.5cm}\\}
\cortext[correspondingauthor]{Corresponding author}
\ead{mostafa.rahimiazghadi@jcu.edu.au}

\begin{abstract}
Memristive devices have shown great promise to facilitate the acceleration and improve the power efficiency of \ac{DL} systems. Crossbar architectures constructed using these \ac{RRAM} devices can be used to efficiently implement various in-memory computing operations, such as \ac{MAC} and unrolled-convolutions, which are used extensively in \acp{DNN} and \acp{CNN}. However, memristive devices face concerns of aging and non-idealities, which limit the accuracy, reliability, and robustness of \acp{MDLS}, that should be considered prior to circuit-level realization. This \ac{OSP} presents \textit{MemTorch}, an open-source\footnote{\url{https://github.com/coreylammie/MemTorch}}  framework for customized large-scale memristive \ac{DL} simulations, with a refined focus on the co-simulation of device non-idealities. MemTorch also facilitates co-modelling of key crossbar peripheral circuitry. MemTorch adopts a modernized software engineering methodology and integrates directly with the well-known PyTorch \ac{ML} library. 
\end{abstract}

\end{frontmatter}

\section{Introduction}
\lettrine{M}{emristive} crossbar architectures~\cite{6709674} have been used to reduce the time complexity of \acp{VMM} used in \acp{DNN} from $\mathcal{O}(n^2)$ to $\mathcal{O}(n)$, and in extreme cases to $\mathcal{O}(1)$~\cite{2019arXiv191005920L}, facilitating the acceleration and improving the power efficiency of \ac{DL} systems~\cite{RahimiAzghadi2020}. However, memristors are still considered an emerging technology, where their reliable manufacturing processes are yet to be achieved. As a result, \ac{DL} architectures realized using memristor crossbars are putative to be prone to severe errors due to a number of device limitations including: finite discrete conductance states, device I/V non-linearity, failure, aging, cycle-to-cycle and device-to-device variability~\cite{Mittal2018,Adam2018}. Consequently, significant research efforts are being made to improve the reliability and robustness of memristive, or \ac{RRAM} crossbars, used to perform \textit{in-situ} learning~\cite{7920854,7966300,Lammie2021c,8468181} and inference~\cite{2019arXiv191005920L,7920854,Mehonic2019,DBLP:journals/corr/abs-1901-10351,Jeong2018,Tsai2018} in \ac{DL} systems. 
A general cross-platform, heterogeneous, high-level, customizable and open-source simulation framework with a refined focus on the co-simulation of device non-idealities could be used to conveniently build, rapidly prototype, and investigate device non-idealities in customized large-scale \acp{MDNN} and \acp{MDLS}. In this \ac{OSP}, we present such a framework, entitled \textit{MemTorch}, for deep memristive learning using crossbar architectures. MemTorch is an open-source~\cite{corey_lammie_2020_3760696} simulation framework that integrates directly with the open-source PyTorch \ac{ML} library that:

\begin{enumerate}
\item Facilitates the cross-platform development and distribution of large-scale passive \ac{0T1R} and active \ac{1T1R} memristive \ac{DL} systems;
\item Places a large emphasis on modeling non-ideal, but inevitable, device characteristics in arbitrary and customizable device models;
\item Supports heterogeneous platforms such as \acp{CPU} and \acp{GPU}; 
\item Has a high-level \ac{API}, which is able to abstract performance-critical tasks described in various low-level languages.
\end{enumerate}

\section{Related Work}
We compare MemTorch to other memristor-based \ac{DNN} frameworks and inference accelerators, which are software-based and do not rely on SPICE modeling, in Table~\ref{lit_review}. More exhaustive comparisons are performed in~\cite{Lammie2022}. Software-based frameworks and inference accelerators use a combination of programming languages to simulate the behavior of memristive devices. Among previous works, DNN+NeuroSim~\cite{Peng2019,Peng2020} and the IBM Analog Hardware Acceleration Kit~\cite{Rasch2021} are the most similar offerings, which integrate with both PyTorch and/or Tensorflow, and can be used to account for non-ideal device characteristics. However, they are largely concerned with algorithm-to-hardware mapping, and are designed to evaluate training and inference accuracy with hardware constraints. They are not designed to model any arbitrary device non-idealities for any behavioral device model. MemTorch, on the other hand, emphasizes the co-simulation of non-ideal device characteristics and generic behavioral device models with stochastic parameters for higher flexibility to simply account for process variance.

\begin{table*}[!t]
\caption{Comparison of MemTorch to other memristor-based DNN simulation frameworks and inference accelerators.$^*$Does not support GPU-accelerated inference and/or parameter mapping.$^\dagger$Models are shared using Google Drive without Application Programming Interfaces (APIs).}
\begin{adjustbox}{width=1\textwidth}
\centering
\begin{tabular}{lcccr}
	\toprule
	\textbf{Simulation framework} & \textbf{Open-source} & \textbf{GPU} & \textbf{Pretrained DNN conversion} & \textbf{Programming language(s)}\\
	\midrule
	RAPIDNN~\cite{DBLP:journals/corr/abs-1806-05794} &  & \phantom{0}\ding{51}$^*$ & \ding{51} & C++ \\
	MNSIM~\cite{7984877} &  &  & \ding{51} & Not Specified \\
	PUMA~\cite{DBLP:journals/corr/abs-1901-10351} &  &  & \ding{51} & C++ \\
	DL-RSIM~\cite{8587661} &  & \ding{51} & \ding{51} & Python \\
	PipeLayer~\cite{7920854} & &  \phantom{0}\ding{51}$^*$ & \ding{51} & C++ \\
	Tiny but Accurate~\cite{2019arXiv190810017M} & \phantom{0}\ding{51}$^\dagger$ &  & \ding{51} & MATLAB \\
	Ultra-Efficient Memristor-Based DNN Framework~\cite{2019arXiv190811691Y} & \phantom{0}\ding{51}$^\dagger$ &  & \ding{51} & C++, MATLAB\\
	Non-ideal Resistive Synaptic Device Characteristic Simulation Framework~\cite{8787884} & & \ding{51} & \ding{51} & Python \\
	\midrule
	Neurosim~\cite{Chen2018a}, NeuroSim+~\cite{Chen2017}, and DNN+NeuroSim~\cite{Peng2019,Peng2020} &  \ding{51} & \ding{51} & \ding{51} &  C++, Python\\
	IBM Analog Hardware Acceleration Kit~\cite{Rasch2021} & \ding{51} & \ding{51} & \ding{51} & Python, C++, CUDA\\
	\textbf{MemTorch} & \ding{51} & \ding{51} & \ding{51} & \textbf{Python, C++, CUDA}\\
	\bottomrule
\end{tabular}
\end{adjustbox}\label{lit_review}
\end{table*}

\section{Software Framework}
The MemTorch simulation framework is programmed in C++, CUDA and Python, with a Python interface. Performance critical tasks are performed using either C++ or CUDA, for \ac{CPU} or \ac{GPU} execution, respectively; otherwise Python is used. MemTorch relies heavily on the open source PyTorch~\cite{paszke2017automatic} \ac{ML} framework, and uses the C++ and Python PyTorch \acp{API} extensively to abstract low-level operations. Consequently, it supports native CPU and GPU operations.

\subsection{Software Architecture}
MemTorch is made up of seven distinct sub-modules. General utility functions, such as data loaders or generic functions, are grouped within \texttt{memtorch.utils}. The \texttt{memtorch.bh} sub-module encapsulates all crossbar models, crossbar mapping and programming methods, crossbar tile mapping and programming methods, memristor models, memristor model window functions, models for all non-ideal device characteristics, quantization methods, and methods to generate stochastic parameters. The \texttt{memtorch.mn} sub-module mimics \texttt{torch.nn} and defines equivalent memristive\texttt{torch.nn.Module} layers. \texttt{memtorch.mn} currently extends \texttt{torch.nn.Linear}, \texttt{torch.nn.Conv1d}, \texttt{torch.nn.Conv2d}, and \texttt{torch.nn.Conv3d}. \texttt{memtorch.mn.Module.patch\_model} can be used to either instantiate new layers, or to patch existing instances. \texttt{memtorch.mn.Module.patch\_model()} iterates through and patches all named modules within classes extending from \texttt{torch.nn.Module} and adds a \texttt{self.tune\_()} method, in addition to other helper methods, to the class instance of the model that automatically patches each selected named module.

The \texttt{memtorch.cpp} sub-module encapsulates all Python-wrapped C++ extensions, whereas the \texttt{memtorch.cu} sub-module encapsulates all Python-wrapped CUDA extensions. Currently, MemTorch uses C++ and CUDA bindings to perform inference for both active and passive modular tiled architectures, and to parallelize quantization operations. The use of bindings can be disabled, and legacy python methods (developed in previous versions of MemTorch) can be used instead using the \texttt{use\_bindings} argument, when patching \texttt{torch.nn.Module} instances.
\texttt{memtorch.examples} sub-module encapsulates general-usage examples and supporting scripts. The \texttt{memtorch.map} sub-module encapsulates all mapping and tuning algorithms used when programming and tuning memristive crossbar arrays. Finally, the \texttt{memtorch.submodule} sub-module encapsulates all external git sub-modules that MemTorch uses. 
We review and present the algorithms and models that are currently built into MemTorch in Appendix A, and our approach to modeling non-ideal device characteristics in Appendix B.

\begin{figure*}[!t]
	\center
	\includegraphics[width=1\textwidth]{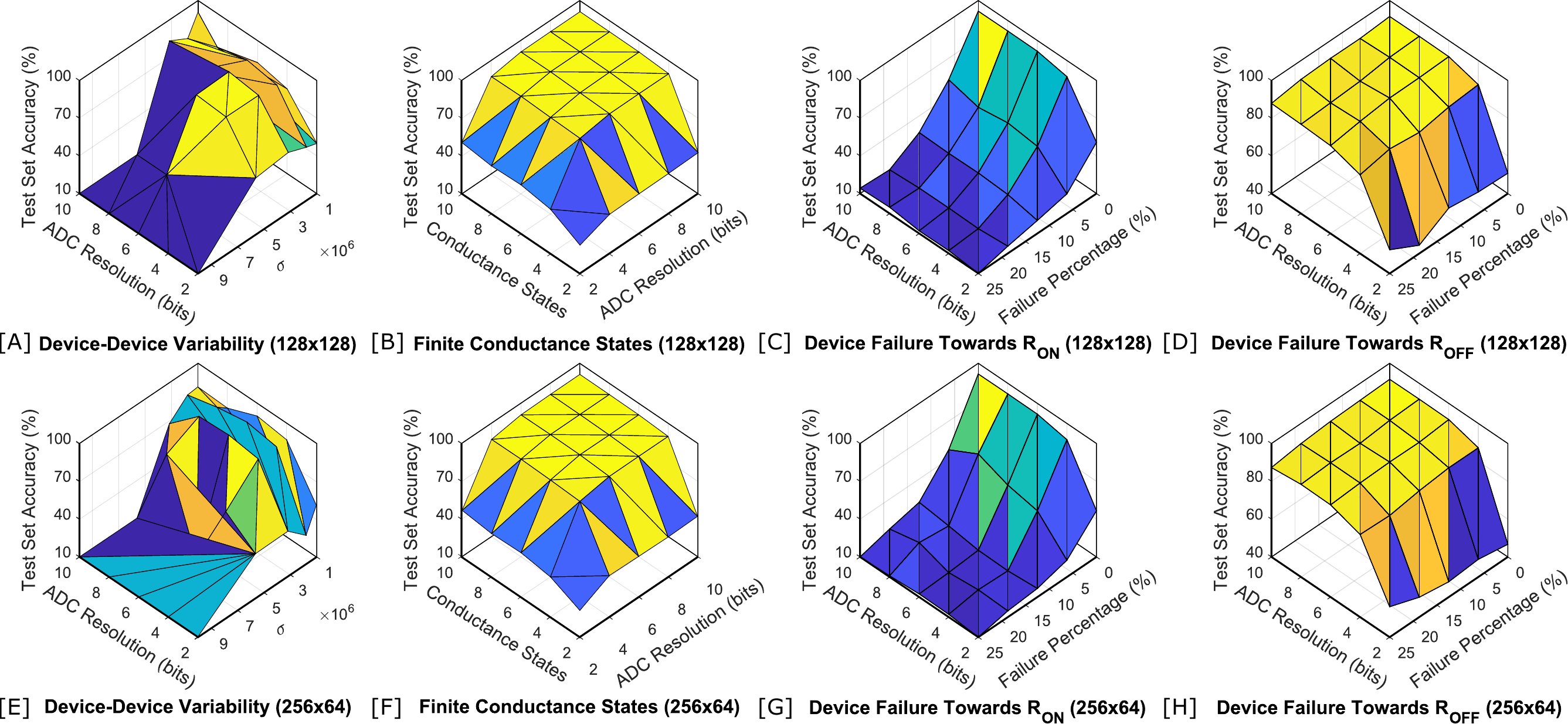}
	\caption{Simulation results when exemplar device non-idealities are considered for classifying CIFAR-10 dataset for two different modular crossbar tile sizes, 128x128 and 256x64. While MemTorch can be used to simulate both passive and active architectures, for demonstration purposes, in this figure, only active architectures are considered.}
	\label{examplar_simulations}
\end{figure*}

\subsection{Software Functionalities}
Complete examples demonstrating the functionality of MemTorch are publicly accessible\footnote{\url{https://github.com/coreylammie/MemTorch/tree/master/memtorch/examples}}. ReadTheDocs\footnote{\url{https://memtorch.readthedocs.io/en/latest/}} is used to explain all functionalities.

\section{Implementation and Empirical Results}
In Fig. \ref{examplar_simulations}, we present exemplar large-scale deep learning simulations to investigate the performance degradation due to device-device variability, finite number of conductance states, and device failure when two different modular crossbar sizes are considered. A separate case study is not presented, as we have previously used MemTorch in other works to perform hand gesture classification~\cite{Azghadi2020}, epileptic seizure prediction~\cite{Lammie2021}, to develop an empirical metal-oxide device endurance and retention model~\cite{Lammie2021b}, and to develop an extended \ac{DSE} methodology for \ac{RRAM} architectures~\cite{Lammie2022b}. Prior to simulation, all convolutional and linear layers from a pre-trained MobileNetV2 for CIFAR-10 were converted to memristive equivalent layers, as explained in detail in~\footnote{\url{https://github.com/coreylammie/MemTorch/tree/master/memtorch/examples/Exemplar_Simulations.ipynb}}, and documented in Appendix C.

\begin{figure*}[!t]
\center
\includegraphics[width=1.0\textwidth]{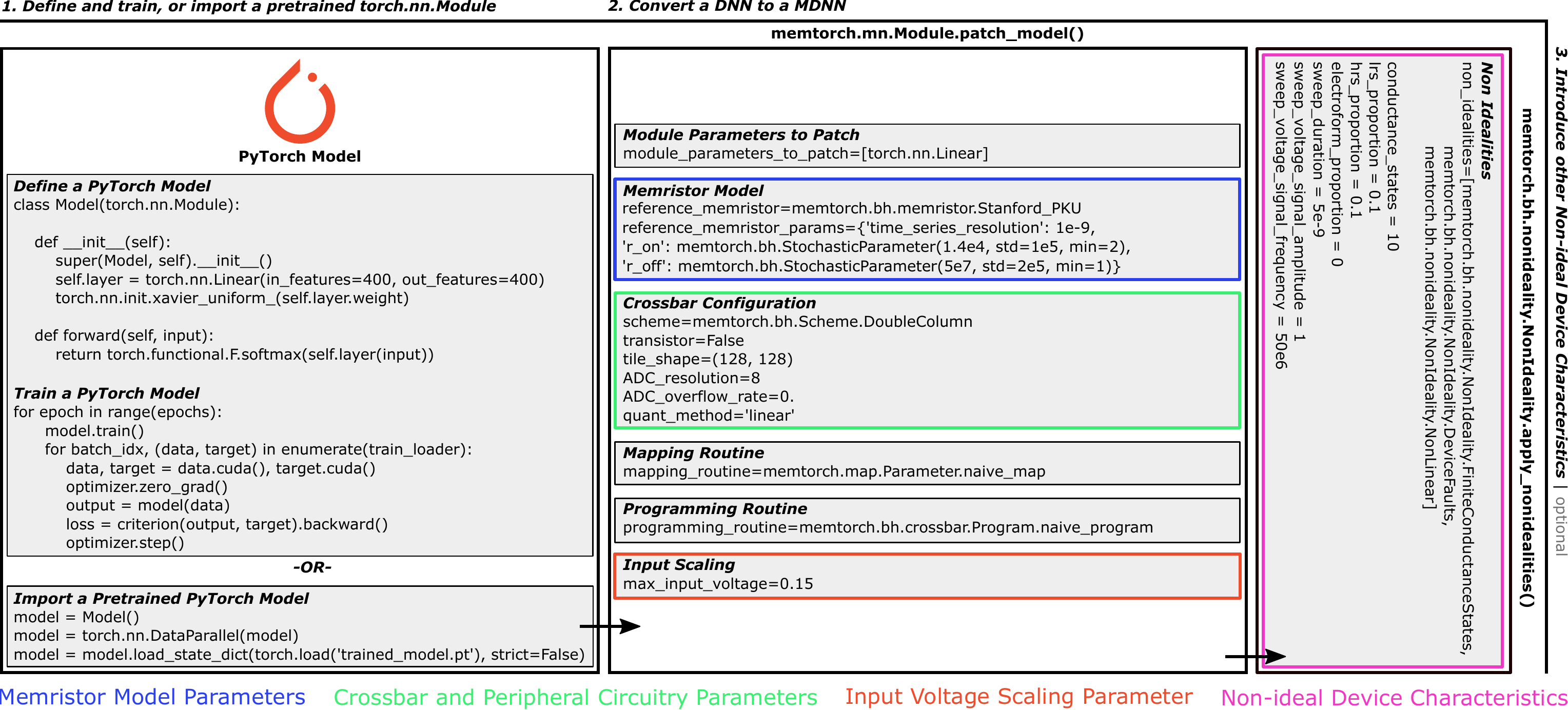}
\caption{Illustration of a typical use-case workflow in MemTorch.} 
\label{overview}
\end{figure*}

\section{Illustrative Example}
To demonstrate MemTorch's intuitive design, we depict a typical use-case work flow in Fig.~\ref{overview}. Here, \texttt{torch.nn.Linear} layers are converted to equivalent memristive layers constructed using modular crossbar tiles, that each contain (128 $\times$ 128) devices, which represent weights using a double-column parameter representation scheme. Inputs are scaled between $\pm0.15V$, and 8-bit \acp{ADC} are used to read out column currents. The Stanford PKU \ac{RRAM} model~\cite{Jiang2014} is used to model TiN/Hf(Al)O/Hf/TiN devices from~\cite{Fantini2014}. Three other non ideal device characteristics were also accounted for including a finite number (10) of discrete conductance states, device faults, and non-linear I/V device behavior.

\section{Conclusion}
We presented an open-source simulation framework, entitled \textit{MemTorch}, for large-scale deep memristive crossbar architectures. We showed that MemTorch is designed with a focus to integrate any desired behavioral or experimental device model, and introduce arbitrary device non-idealities, while co-simulating crossbar and peripheral circuitry. We compared MemTorch to similar works, detailed its package structure, and performed exemplar simulations to demonstrate its functionality. We hope that MemTorch will be adopted and expanded by the community to advance memristive deep learning research and development endeavours.  

\newpage
\section*{Acknowledgements}
CL acknowledges the JCU DRTPS, CAS Society Pre-Doctoral Grant, and the IBM PhD Fellowship. MRA acknowledges a JCU Rising Star ECR Leadership Grant.

\appendix
\section{Algorithms and Models}
This appendix reviews and presents the algorithms and models that are currently built into MemTorch.

\subsection{Memristive Device Models}
Within MemTorch, we use five base memristive device models that extend the \texttt{memtorch.bh.Memristor.Memristor} base class for our simulations. These include the linear ion drift model by~\cite{Strukov2008}; the VTEAM model by~\cite{7110565}, which is a general model for voltage-controlled memristors that can be used to fit a large range of experimental device data; the Stanford PKU \ac{RRAM} model~\cite{Jiang2014}, which describes switching performance for bipolar metal oxide \ac{RRAM}; and two versions of the data-driven Verilog-A \ac{RRAM} model~\cite{Messaris2018}, which expresses device current-voltage characteristics and resistive switching rate as a function of the bias voltage and the initial resistive state of each device. For each base model, finite differences is used to obtain a numerical solution for each discretized time-step, $dt$. While only five base memristive models are currently supported natively, others, which can model the equivalent conductance of a memristive device for an arbitrary applied voltage signal, such as those modelling \ac{PCM} or other device technology behavior, can easily be integrated modularly by extending \texttt{memtorch.bh.Memristor.Memristor}. 

\begin{figure*}[!t]
	\center
	\includegraphics[width=1.0\textwidth]{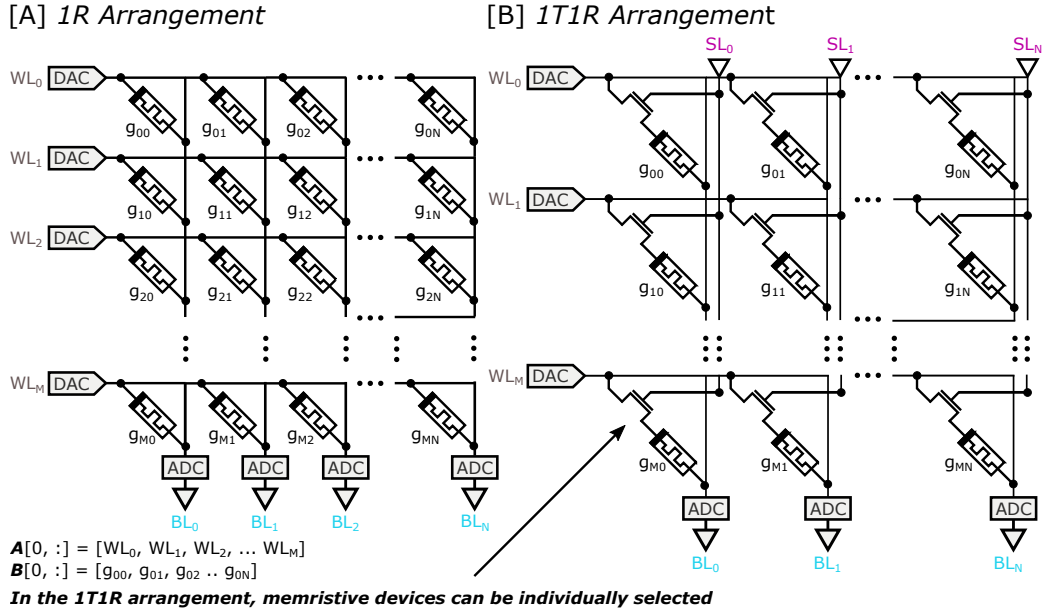}
	\caption{Depiction of an $M \times N$ [A] 1R (0T1R) crossbar architecture and a [B] 1T1R crossbar architecture. Matrix-vector and matrix-matrix multiplication can be performed by encoding and presenting a scaled input vector or matrix $\bm{A}$ as voltage signals to each row of the crossbar's \acp{WL}. As shown in [A], assuming a linear I/V relationship, the total current in each column's \ac{BL} is linearly proportional with the sum of the multiplication of the \ac{WL} voltages and conductance values in that column, i.e., $\bm{BL}[0, :] \propto \bm{A}$[0, :] $\times$ $\bm{B}$. In the 1T1R arrangement [B], individual memristive devices can be selected using \acp{SL}.}
	\label{fig:crossbar_arrangement}
\end{figure*}

\subsection{Window Functions}
Within memristive device models, window functions are widely employed to restrict the changes of the internal state variables to specified intervals~\cite{DBLP:journals/corr/abs-1811-06649}. MemTorch currently natively supports the Biolek~\cite{Biolek_spicemodel}, Jogelkar~\cite{Joglekar2009}, and Prodromakis~\cite{5934403} window functions, and can easily be extended to support others.

\subsection{Memristive Crossbar Architectures}
Memristive devices can be arranged within crossbar architectures to perform \acp{VMM}, which are used extensively in forward and backward propagations within \acp{DNN}. There are two commonly used crossbar architecture configurations, namely 1-Transistor 1-Resistor (1T1R), and 1-Resistor (1R or /\ac{0T1R}), which are both depicted in Fig.~\ref{fig:crossbar_arrangement}. In 1T1R arrangements, one transistor is used to select and control each memristive device, whereas in 1R arrangements, rows and columns of memristive devices are positioned perpendicular to each other, with memristive devices sandwiched in-between.

The product of a vector and a matrix or, in a more general form, two matrices, $\bm{A}$ of size $(M \times C)$ and $\bm{B}$ of size $(C \times N)$, can be computed using a crossbar-architecture, as illustrated in Fig. \ref{fig:crossbar_arrangement}, where $\bm{A}$ represents scaled input voltage signals and $\bm{B}$ is encoded within the crossbar as memristor conductances. Separate \acp{ADC} can be used to read out the current of each column in parallel, as depicted, or sample and hold circuits can be used in conjunction with a single \ac{ADC} per crossbar, that can be used to read out the current of each column sequentially using \ac{TDM}. As the output current of each column is linearly proportional to the elements of $\bm{A} \bm{B}$, a linear constant, $K$, is used to correlate the \ac{ADC} readout of each column accordingly. By separately presenting each row of $\bm{A}$ to the crossbar through \acp{WL}, all rows of $\bm{A} \bm{B}$ can be computed.

Because memristors cannot be programmed to have negative conductances, within \acp{MDNN}, weight matrices can either be represented using two devices per weight~\cite{Alibart2013}, as described by (\ref{ab_two}), 
\begin{equation}\label{ab_two}
	\bm{A} \bm{B} = K \sum_{i=0}^C \bm{A}[i, :] (g_{\text{pos}}[i, j] - g_{\text{neg}}[i, j]), \textnormal{for $j=0$ to $N$},
\end{equation}
	or using a single device per weight~\cite{truong2014new,2019arXiv190609395L} using complex weight mapping algorithms or current mirrors, as described by (\ref{ab_one}),
\begin{equation}\label{ab_one}
	\bm{A} \bm{B} = K \sum_{i=0}^C \bm{A}[i, :] (g[i, j] - g_{m}), \textnormal{for $j=0$ to $N$}.
\end{equation}
\noindent For the single-column case, the current through $g_m$, used to mirror a current $-2V / (\bar{R}_{\text{ON}} + \bar{R}_{\text{OFF}})$ to each crossbar, is copied to each column and subtracted from all memristor columns. This current can be realized using a diode-connected NMOSFET by adjusting the NMOSFET channel width so that it has a passive resistance $g_m$. From this stage forward, we refer to the weight matrix representation methodology adopted, that is, weather two devices are used to encode each weight, i.e, differential weight mapping, one device is used to encode each weight, or another configuration is used to encode weight matrices, as the parameter representation scheme.

\subsubsection{Modular memristive crossbar tiles}
Mapping complete unrolled neural network layers into large memristive crossbar architectures often results in poor performance. This is due to non-ideal device characteristics that introduce substantial current variability when accumulated currents from columns with a large number of devices are read out. When one or two large crossbars are used, for single-column and double-column parameter representation schemes, respectively, they cannot easily be modularized because customized crossbar shapes are required to represent each individual layer. Instead of using large crossbars, modular crossbar tiles~\cite{Mountain2018} can be used that map layers into multiple uniformly sized crossbars, commonly referred to in literature as \textit{crossbar tiles}. 

One large crossbar of size ($M \times N$) can be mapped using ceil($M/S_0$)$\times$\\ceil($N/S_1$) crossbar tiles, each with a size of ($S_0 \times S_1$), where the total utilization, $\rho$, of all crossbar tiles can be determined using (\ref{eq:utilization}),
\begin{equation}\label{eq:utilization}
    \rho = \frac{MN}{\text{ceil}(M/S_0)\text{ceil}(N/S_1)S_0S_1}.
\end{equation}

\noindent Duplication of crossbar tiles and \ac{TDM} can be used to regulate mapping to improve the array utilization and computation time by balancing latency among layers~\cite{Wang2019}.

\subsubsection{Memristor crossbar programming}
The conductance of memristive devices can be altered between a low resistance state $R_{\text{ON}}$ and a high resistance state $R_{\text{OFF}}$, by applying programming voltage pulses with different intervals and amplitudes. While individual devices within crossbars can be selected and programmed within 1T1R cells, in 1R arrangements, when a voltage is applied to a specific device, a non-zero voltage (usually half that of the nominal programming pulse amplitude) is applied to all other devices in the same row and column. Consequently, various multistage programming~\cite{6069931,7147235,7180590,5482157} and corrective methods~\cite{8326998,2019arXiv191005920L,6709674}, which can use analog voltage wave-forms, are often used to ensure the difference between the programmed conductance states and the conductance states-to-program are within an acceptable tolerance.
	
\begin{algorithm}[!t]
	\caption{Memristor crossbar programming algorithm.}
	\begin{algorithmic}
		\renewcommand{\algorithmicrequire}{\textbf{Input:}}
		\renewcommand{\algorithmicensure}{\textbf{Output:}}
		\REQUIRE Array containing all continuous weights in a given layer, $\bf{w}$, HRS/LRS ratio, $p_L$.
		\ENSURE Equivalent memristive crossbars conductance values, $\bf{g}$, indexed using $i$ and $j$.
		\STATE $\bf{w}$ = abs($\bf{w}$)
		\STATE $\bf{w}$ = descending\_order($\bf{w}$)
		\STATE $s$ = size($\bf{w}$)
		\STATE index = int($p_L \cdot s$)\
		\STATE $\bf{w}_{\text{max}} = \bf{w}$[index]
		\STATE $\bf{w}_{\text{min}} = \bf{w}_{\text{max}} / (R_{\textnormal{OFF}} / R_{\textnormal{ON}})$
		\STATE $\bf{w} = \text{clip}(\bf{w}, \bf{w}_{\text{min}}, \bf{w}_{\text{max}})$
		\STATE $\bm{g}[i, j] = \frac{(R_{\textnormal{ON}} - R_{\textnormal{OFF}}) \cdot (\sigma(\bm{w})[i, j] - \bf{w}_{\text{min})}} {|\bm{w}|_{\textnormal{max}} - \bf{w}_{\text{min}}} + R_{\textnormal{OFF}}$
	\end{algorithmic}\label{prog_train_alg}
\end{algorithm}

\subsubsection{Memristor crossbar tuning}
The total current of each column in an ideal memristive crossbar is linearly proportional to the output elements of the \ac{VMM} resultant vector. Consequently, after each \ac{DNN} layer's weights are programmed into a crossbar or group of tiles, linear regression can be used to correlate the output current of each column with any desired output to determine $K$ for the crossbar or group of tiles, given a randomly generated input matrix that is sufficiently large. On account of device-device variations and device failures, further tuning is often required to recover accuracy loss and mitigate variances between intended and actual device conductance values. Tuning methods can either be used pre-programming~\cite{Zhang:2019:HSM:3287624.3287707}, to improve robustness and reduce susceptibility to error, or post-programming by retraining device-specific conductance values~\cite{2019arXiv190810017M}.

\subsubsection{Memristor crossbar weight mapping}
Weights, denoted using $\bm{w}$, within unrolled convolutional layers~\cite{Chellapilla2006HighPC} and linear layers can be mapped to equivalent conductance values, $\bm{g}$, using (\ref{simple_map}). 
\begin{equation}\label{simple_map}
	\bm{g}[i, j] = \frac{(g_{\textnormal{ON}} - g_{\textnormal{OFF}}) (\sigma(\bm{w})[i, j] - \bf{w}_{\text{min})}} {|\bm{w}|_{\textnormal{max}} - \bf{w}_{\text{min}}} + g_{\textnormal{OFF}},
\end{equation}
where $\bf{w}_{\text{min}}$ represents the minimum weight value to encode, and $\bf{w}_{\text{max}}$ represents the maximum weight value to encode. $g_{\textnormal{ON}} = 1 / R_{\textnormal{ON}}$ and $g_{\textnormal{OFF}} = 1 / R_{\textnormal{OFF}}$. When two crossbars are used to represent weight, crossbars containing positive components will have $\sigma(\bm{w}) = \bm{w}[\bm{w} \geq 0]$, while crossbars containing negative components will have $\sigma(\bm{w}) = \bm{w}[\bm{w} \leq 0]$. When a single crossbar is used to represent weights, $\sigma(\bm{w}) = \bm{w} - g_m$.
	
To reduce the inner weight gap in a given device, Algorithm~\ref{prog_train_alg} in \cite{Mehonic2019} can be used to exclude a small proportion, $p_L$, of weights with the absolute largest values to reduce the variability effect of non-ideal memristive devices.

\subsubsection{Passive memristor crossbar architectures}
When modeling passive memristor crossbar architectures, source and line resistances should be accounted for. MemTorch utilizes a comprehensive crossbar array model with solutions for source and line resistances, as described in~\cite{Chen2013}, to solve for node voltages $\bm{V}$, within passive crossbar architectures in each simulation time-step. Specifically, linear matrix algebra is used to solve~(\ref{eq:passive})
\begin{equation}\label{eq:passive}
    \left[ {\begin{array}{cc}
    \bm{A} & \bm{B} \\
    \bm{C} & \bm{D} \\
  \end{array} } \right] \bm{V} = \bm{E},
\end{equation}

\noindent where for a crossbar of size ($M \times N$), $\bm{A}$, $\bm{B}$, $\bm{C}$, $\bm{D}$ are all ($MN \times MN$) matrices, and $\bm{E}$ is a $(2MN \times 1)$ vector. All matrices and vectors in~(\ref{eq:passive}) are derived and defined in~\cite{Chen2013}. As the concatenated $\bm{ABCD}$ matrix is sparse, traditional linear matrix algebra methods cannot be easily used to solve~(\ref{eq:passive}), because they require a prohibitive amount of memory. Instead, sparse supernodal LU factorization with partial pivoting for general matrices~\cite{Li2011a} is used, as it is parallelizable, and has demonstrated the best empirical numerical performance, compared to related techniques, e.g., QR decomposition~\cite{Matstoms1994}.

\subsection{C++ and CUDA Bindings}
Numerous performance critical operations, including tiled \acp{VMM}, linear matrix algebra, and quantization, are accelerated using C++ and CUDA bindings. \color{black}\texttt{PYBIND11\_MODULE()} is a method within the pybind11~\footnote{\url{https://github.com/pybind/pybind11}} python library~\cite{pybind11} that exposes C++ types in Python to enable seamless operability between C++11, CUDA, and Python. This library is used within MemTorch to overload method pointers and to expose C++ and CUDA functions to the developed Python \ac{API}. The Eigen~\cite{eigenweb} C++ template library is used extensively to perform various linear algebra operations, as many Eigen functions can be compiled for use within CUDA kernels using \texttt{\_\_device\_\_ \_\_host\_\_} function type qualifiers.

\section{Modeling Non-Ideal Device Characteristics} 
\begin{figure*}[!t]
	\center
	\includegraphics[width=1\textwidth]{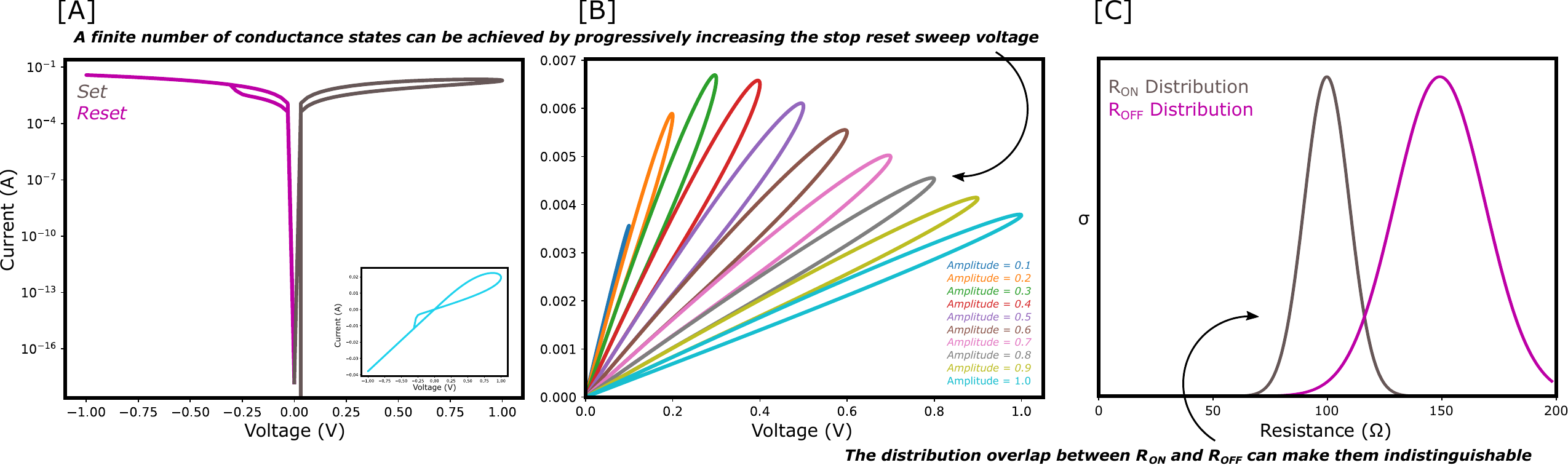}
	\caption{Depiction of [A] device I/V characteristics, and [B] reset voltage double-sweeps demonstrating gradual switching from $R_{\text{ON}}$ to $R_{\text{OFF}}$, which can be used to achieve 10 finite stable conductance states for the VTEAM model using the TEAM~\cite{6353604} model's parameters, with a linear dependence on $w$, achieved using sinusoidal signals with a fixed frequency of 50 MHz. [C] shows distributions of $R_{\text{ON}}$ and $R_{\text{OFF}}$, which are caused by device-device variability, for a memristive device with $\bar{R}_{\text{ON}} = 100\Omega$ and $\bar{R}_{\text{OFF}} = 150\Omega$. In [C], overlapped regions are indistinguishable from each other.}
	\label{non_ideal_fig}
\end{figure*}

Non-ideal device characteristics can either be encapsulated within device-specific memristive models, or introduced to base (generic) models using the \texttt{memtorch.bh.nonideality} sub-module. This sub-module can currently be used to introduce four non-ideal device characteristics to memristive device models: device-device variability, finite number of discrete conductance states, device failure, and non-linear I/V device characteristics. We leave native support for modeling other non-ideal device characteristics to future releases. Three of the non-ideal device characteristics that are currently supported by MemTorch are shown in Fig. \ref{non_ideal_fig}. Fig. \ref{non_ideal_fig}[A] depicts typical non-linear I/V device characteristics using a set-reset curve and an inset hysteresis loop. Fig.~\ref{non_ideal_fig}[B] demonstrates gradual switching, which is used to achieve a finite number of stable conductance states, and Fig. \ref{non_ideal_fig}[C] shows overlapping distributions of $R_{\text{ON}}$ and $R_{\text{OFF}}$, which is caused by device-to-device variability.

\subsection{Device-to-device Variability}
Device-to-device variability is modeled stochastically using \\\texttt{memtorch.bh.StochaticParameter}. Stochastic parameters are generated using the \texttt{memtorch.bh.StochaticParameter.StochaticParameter()}\\method, which accepts an arbitrary number of keyword arguments, that are used to sample from a \texttt{torch.distributions} each time a device model is instantiated. To model device-device variability, we use stochastic parameters to sample $R_{\textnormal{ON}}$ and $R_{\textnormal{OFF}}$ from a normal distribution with $\sigma R_{\textnormal{ON}} = \sigma$ and $\sigma R_{\textnormal{OFF}} = 2\sigma$. $\sigma R_{\textnormal{OFF}} > \sigma R_{\textnormal{ON}}$, as the variability of $R_{\textnormal{OFF}}$ has been demonstrated to be larger than $R_{\textnormal{ON}}$~\cite{8702245}. As depicted in Fig. \ref{non_ideal_fig}[C], device-device variability can cause the distribution of $R_{\textnormal{ON}}$ and $R_{\textnormal{OFF}}$ to overlap, resulting in $R_{\textnormal{ON}}$ and $R_{\textnormal{OFF}}$ occupying the same conductance regions.

\subsection{Cycle-to-cycle Variability}
\ac{C2C} variability~\cite{8548606} is modeled stochastically, similarly to device-to-device variability, using stochastic parameters for $R_{\text{ON}}$ and $R_{\text{OFF}}$. \texttt{memtorch.bh.nonideality.DeviceFaults.apply\_cycle\_variability()} is used to sample $R_{\textnormal{ON}}$ and $R_{\textnormal{OFF}}$ from a normal distribution with $\sigma R_{\textnormal{ON}} = \sigma$ and $\sigma R_{\textnormal{OFF}} = 2\sigma$ after each SET RESET cycle.
	
\begin{figure*}[!t]
	\center
	\includegraphics[width=1\textwidth]{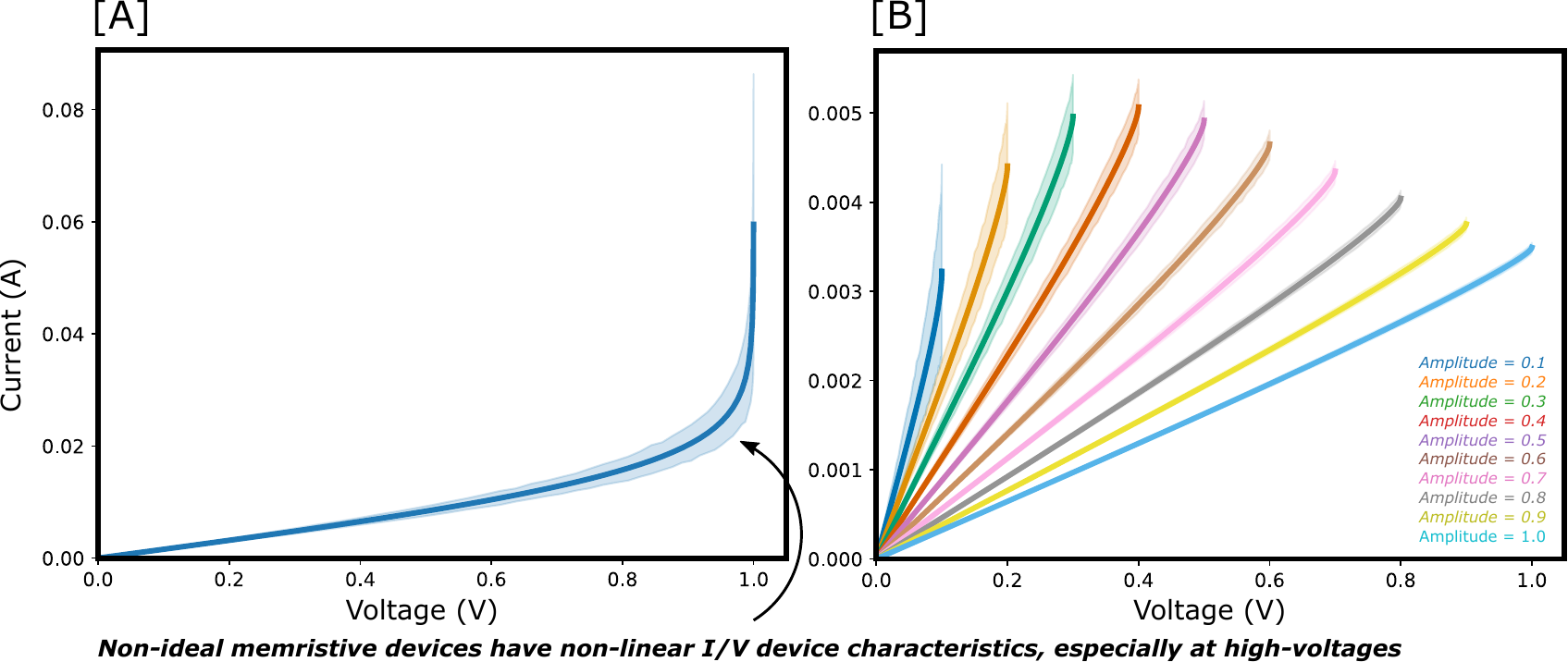}
	\caption{Non-linear I/V characteristics for 100 devices (instances) of the VTEAM model using the TEAM~\cite{6353604} model's parameters, with a linear dependence on $w$, achieved using sinusoidal signals with a fixed frequency of 50 MHz. $R_{\text{ON}}$ and $R_{\text{OFF}}$ were stochastically sampled from a normal distribution with $\bar{x} = 50, \sigma = 25$, and $\bar{x} = 1000, \sigma = 50$, respectively. [A] depicts I/V characteristics for devices with an infinite number of discrete conductance states. [B] depicts I/V characteristics for devices with a finite number of discrete conductance states.}
	\label{placeholder_10}
\end{figure*}

\subsection{Finite Number of Discrete Conductance States}
Realistic memristive devices are non-ideal and have a finite number of stable discrete electrically switchable conductance states, bounded by a low-conductance semiconducting state $R_{\textnormal{OFF}}$, and a high-conductance metallic state, $R_{\textnormal{ON}}$~\cite{Yi2016}. Previous works have investigated evenly spaced conductance or resistance states, and have demonstrated that, assuming they are relatively uniformly distributed, the spacing between states is not critical~\cite{Mehonic2019}. 
	
Therefore, deterministic discretization~\cite{8267253} can be used to represent a finite number of electrically switchable conductance states, as depicted in Fig. \ref{non_ideal_fig}[B]. In order to efficiently quantize a tensor to a defined finite number of quantization states, in which each element can have a different range, CUDA kernels are used to perform a binary search on sorted tensors (generated using the \texttt{linspace} algorithm in C++) containing defined quantization states in $\mathcal{O}(n \textnormal{log}(n))$, where $n$ is the number of quantized states.

\subsection{Device Failure}
Memristive devices are susceptible to failure, by either failing to eletroform at a pristine state, or becoming stuck at high or low resistance states~\cite{Mehonic2019}. MemTorch incorporates a specific function for accounting for device failure in simulating \ac{DL} systems. Given a \texttt{nn.Module}, \texttt{memtorch.bh.nonideality.\\DeviceFaults.apply\_device\_faults()} sets the conductance of a proportion of devices within each crossbar to $R_{\text{ON}}$ or $R_{\text{OFF}}$. It is assumed that the total proportion of devices set to $R_{\text{OFF}}$ is equal to the proportion of devices that fail to eletroform at pristine states plus the proportion of devices stuck at a high resistance state. However, these proportions and the ratio of device failures can be manipulated as desired. Devices are chosen at random using \texttt{np.random.choice()}.

\subsection{Non-linear I/V Characteristics}
Non-ideal memristive devices have non-linear I/V device characteristics,  especially at high voltages, which are difficult to accurately and efficiently model~\cite{Mehonic2019}. We demonstrate these characteristics using Fig.~\ref{placeholder_10}[A], by depicting the I/V curve of the VTEAM model between 0--1V using the TEAM~\cite{6353604} model's parameters. The \texttt{memtorch.bh.nonideality.NonLinear\\.apply\_non\_linear()} method can be used to efficiently model non-linear device I/V characteristics during inference for devices with an infinite number of discrete conductance states, and for devices with a finite number of conductance states. For cases where devices are not simulated using their internal dynamics, it is assumed that the change in conductance during read cycles is negligible.

\subsubsection{Devices with an infinite number of discrete conductance states}
The \texttt{memtorch.bh.nonideality.NonLinear.apply\_non\_linear()}\\method uses two methods to efficiently model non-linear device I/V characteristics for devices with an infinite number of discrete conductance states during inference:
	
\begin{enumerate}
	\item During inference, each device is simulated for a single timestep, \\\texttt{device.time\_series\_resolution}, using \texttt{device.simulate()}.
	\item Post weight mapping and programming, the I/V characteristics of each device are determined using a single reset voltage sweep. The I/V characteristics of each device are stored, and used as \acp{LUT} to compute device output currents during inference. 
\end{enumerate}

\subsubsection{Devices with a finite number of discrete conductance states}
The \texttt{memtorch.bh.nonideality.NonLinear.apply\_non\_linear()}\\method effectively models non-linear I/V characteristics for devices with a finite number of discrete conductance states by determining the I/V characteristics of each device post weight mapping and programming during several single reset voltage sweeps. Fig.~\ref{placeholder_10}[B] depicts sweeps for 100 stochastic devices with 10 finite discrete conductance states. These are stored and used as \acp{LUT} to compute device output currents during inference, where each I/V curve corresponds to each finite discrete conductance state. In Fig.~\ref{placeholder_10}[B], the smallest voltage amplitude corresponds to the finite conductance state closest to $R_{\text{ON}}$, whereas the largest voltage amplitude corresponds to the finite conductance state closest to $R_{\text{OFF}}$.

\section{Exemplar Simulation Details}
For all simulations performed to obtain the results presented in Fig.~\ref{examplar_simulations}, we followed the following training and test  procedure. We first augmented a pretrained MobileNetV2 \ac{CNN} trained using the CIFAR-10 training set. All convolutional and linear layers within the network were sequenced with batch-normalization layers with fixed affline parameters to normalize outputs. The network was trained until improvement on the validation set was negligible (for 100 epochs) with a batch size of $\Im=256$. The initial learning rate was $\eta = 1e-1$, which was decayed by an order of magnitude every 40 training epochs. \ac{SGD} was used to optimize network parameters and Cross Entropy (CE)~\cite{DBLP:journals/corr/abs-1805-07836} was used to determine network losses. The network achieved $>90\%$ accuracy on the CIFAR-10 test set.

When implementing the \acp{MDNN}, each memristive layer's weights were mapped to a double column line crossbar architecture adopting a 1T1R arrangement. Linear regression was used to correlate the output current of each column and its corresponding output to determine $K$ for each crossbar, given a randomly generated input matrix sampled from a uniform distribution between $\pm1.0$. For linear layers, the random inputs had a size of (8 $\times$ in\_features), while for convolutional layers the random inputs had a size of (8 $\times$ in\_channels $\times$ 32 $\times$ 32). Unless otherwise stated, inputs to memristive layers were scaled from -0.3 to 0.3, to emulate voltage signals between $\pm$0.3V, which were applied to the word-lines of each memristive crossbar. All device models originated from the VTEAM model, with $\bar{R}_{\text{ON}}$ = 1.4e4$\Omega$ and $\bar{R}_{\text{OFF}}$ = 5e7$\Omega$, to model TiN/Hf(Al)O/Hf/TiN devices from~\cite{Fantini2014}. 

Implementations are investigated using modular crossbar tiles of size 128$\times$128 and 256$\times$64, as these have previously been demonstrated to be effective in terms of utilization and power efficiency~\cite{Wang2019}. While power and latency balancing is beyond the scope of MemTorch 1.1.5, 256$\times$64 tile size enables higher operation throughput and more analog operations per ADC compared to 128$\times$128 tile size~\cite{Wang2019}. However, the area utilization may be lower for arrays with more than 64 columns considering the number of output channels.

\bibliography{References.bib}

\clearpage
\section*{Required Metadata}
\label{}

\section*{Current executable software version}

\begin{table}[!h]
\begin{tabular}{|l|p{6.5cm}|p{6.5cm}|}
\hline
\textbf{Nr.} & \textbf{(executable) Software metadata description} & \textbf{Please fill in this column} \\
\hline
S1 & Current software version & 1.1.5\\
\hline
S2 & Permanent link to executables of this version  & \url{https://github.com/coreylammie/MemTorch/releases/tag/v1.1.5}\\
\hline
S3 & Legal Software License & GPLv3\\
\hline
S4 & Computing platform/Operating System & Linux, OS X, Microsoft Windows\\
\hline
S5 & Installation requirements \& dependencies & A working Python interpreter ($\geq$3.6). If CUDA is True in \texttt{setup.py}, CUDA Toolkit ($\geq$10.1) and Microsoft Visual C++ Build Tools are required. If CUDA is False in \texttt{setup.py}, Microsoft Visual C++ Build Tools are required. All Python requirements are listed in \url{https://github.com/coreylammie/MemTorch/blob/master/requirements.txt}\\
\hline
S6 & If available, link to user manual - if formally published include a reference to the publication in the reference list & \url{https://memtorch.readthedocs.io/en/latest/} \\
\hline
S7 & Support email for questions & coreylammie@gmail.com, corey.lammie@jcu.edu.au\\
\hline
\end{tabular}
\caption{Software metadata (optional)}
\label{} 
\end{table}

\newpage
\section*{Current code version}
\label{}

\begin{table}[!h]
\begin{tabular}{|l|p{6.5cm}|p{6.5cm}|}
\hline
\textbf{Nr.} & \textbf{Code metadata description} & \textbf{Please fill in this column} \\
\hline
C1 & Current code version & 1.1.5\\
\hline
C2 & Permanent link to code/repository used of this code version & \url{https://github.com/coreylammie/MemTorch}\\
\hline
C3 & Legal Code License   & GPLv3\\
\hline
C4 & Code versioning system used & git\\
\hline
C5 & Software code languages, tools, and services used & Python, C++, CUDA.\\
\hline
C6 & Compilation requirements, operating environments \& dependencies & A working Python interpreter ($\geq$3.6). If CUDA is True in \texttt{setup.py}, CUDA Toolkit ($\geq$10.1) and Microsoft Visual C++ Build Tools are required. If CUDA is False in \texttt{setup.py}, Microsoft Visual C++ Build Tools are required. All Python requirements are listed in \url{https://github.com/coreylammie/MemTorch/blob/master/requirements.txt}\\
\hline
C7 & If available Link to developer documentation/manual & \url{https://memtorch.readthedocs.io/en/latest/}\\
\hline
C8 & Support email for questions & coreylammie@gmail.com, corey.lammie@jcu.edu.au\\
\hline
\end{tabular}
\caption{Code metadata (mandatory)}
\label{} 
\end{table}

\end{document}